# Secured Wireless Communication using Fuzzy Logic based High Speed Public-Key Cryptography (FLHSPKC)

Arindam Sarkar
Department of Computer Science & Engineering
University of Kalyani
Kalyani-741235, Nadia, West Bengal, India.

J. K. Mandal
Department of Computer Science & Engineering
University of Kalyani
Kalyani-741235, Nadia, West Bengal, India.

*Abstract*— In this paper secured wireless communication using fuzzy logic based high speed public-key cryptography (FLHSPKC) has been proposed by satisfying the major issues likes computational safety, power management and restricted usage of memory in wireless communication. Wireless Sensor Network (WSN) has several major constraints likes' inadequate source of energy, restricted computational potentiality and limited memory. Though conventional Elliptic Curve Cryptography (ECC) which is a sort of public-key cryptography used in wireless communication provides equivalent level of security like other existing public–key algorithm using smaller parameters than other but this traditional ECC does not take care of all these major limitations in WSN. In conventional ECC consider Elliptic curve point $p$, an arbitrary integer $k$ and modulus $m$, ECC carry out scalar multiplication $kP$ mod $m$, which takes about 80% of key computation time on WSN. In this paper proposed FLHSPKC scheme provides some novel strategy including novel soft computing based strategy to speed up scalar multiplication in conventional ECC and which in turn takes shorter computational time and also satisfies power consumption restraint, limited usage of memory without hampering the security level. Performance analysis of the different strategies under FLHSPKC scheme and comparison study with existing conventional ECC methods has been done.

## I. INTRODUCTION

Now-a-days a number of public-key cryptographic algorithms are obtainable for securing data like authentication, non-repudiation, integrity and confidentiality [7, 8, 16]. Among these algorithms a few of them are faster and some of the other algorithms are slower. WSN consist of small nodes that sense their surroundings, process data and communicate through wireless connection. In wired data networks node rely on pre deployed trusted server assist to set up belief connection but in case of WSN, because of limited memory CPU power , and energy of sensors their trusted authorities do not exist [17].

Due to the restraint in the bandwidth, computational potency, power availability or storage in mobile devices, the PKC-based remote authentication schemes are not appropriate for mobile devices.

For these reasons WSN needs faster public-key cryptographic algorithm which satisfied all these constraints [18, 19, 20, 21]. Neal Koblitz and victor Miller was proposed Elliptic Curve Cryptography (ECC) [2]. The advantages of ECC over RSA, DSA, ElGamal [3], Rabin, Diffie-Hellman Key exchange [1] are that it provides efficient security using shorter key and offers

1) *minimum space complexity for key storage.*
2) *less energy cost for performing arithmetic operations.*
3) *minimum time complexity for transmission of keys.*

These features make ECC best substitute to offer security in WSN [15]. ECC offers a popular solution to the problem of implementing public key cryptography on mobile computing devices. The security of RSA, the most popular algorithm in other fields is based on the hardness of integer factorization. ECC is based on the Elliptic Curve Discrete Logarithm Problem (ECDLP) which is the best algorithm for solving ECC takes fully exponential time [15].

Consider base point P, arbitrary integer k, scalar multiplication will be Q = kP. Using ECDLP to find k from P and Q are a hard problem. Whereas the best algorithm for solving RSA, DSA takes sub exponential amount of time [12].

ECC keys are shorter than their RSA analogues, while achieving the similar security level. A 160-bit ECC key is approximately corresponding to a 1024-bit RSA key. So ECC based system is normally more competent and utilize less resources than RSA and hence ECC has appeared as a promising choice to traditional public key techniques on WSNs, because of its lesser processing and storage requirements. ECC is faster than RSA for decryption, but slower than RSA for encryption.

In ECC-based authentication algorithm, Elliptic Curve scalar multiplication is core operation, but this operation is the most time consuming operation. This operation takes 80% of executing time.

Hardware implementation of ECC involves tree-layer hierarchical strategy namely finite field arithmetic, point arithmetic and scalar multiplication as shown in figure 2.

There are many attempts to implement this 3-tier in order to obtain swift computation, reduce power consumption and reduce storage space.

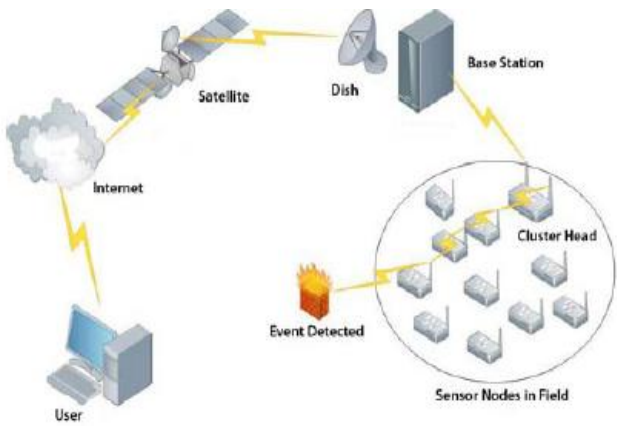

Figure 1. Structural design of WSN

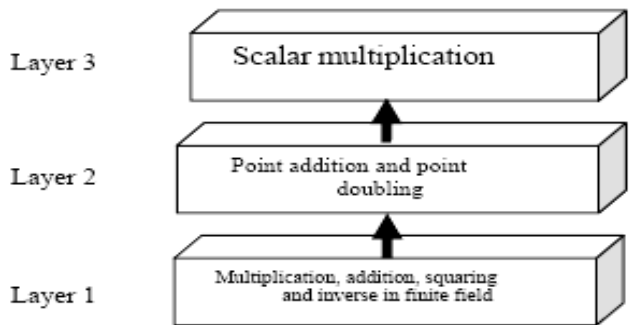

Figure 2. 3-tier architecture of ECC scalar multiplication

In this paper for implementing ECC over binary fields, improving its performance and viability an optimized technique has been proposed. The most important operations in any ECC based techniques such as key exchange or encryption is the scalar multiplication. This point scalar multiplication is achieved by repeated point addition and doubling.

In this paper FLHSPKC scheme has been proposed for scalar multiplication which outstandingly improves the computational competence of scalar multiplication. In proposed FLHSPKC scheme 4 techniques has been proposed to extend the performance of classical ECC by providing following features and comparisons among these newly proposed scheme has also been done.

- Requirement of less amount of memory during execution
- Does not store individual operation at some point that means it do not store pre computed values but it keeps full power computation.
- It is as fast as the traditional ECC.

## II. RELATED WORK

Gura *et al.* [14] showed that scalar multiplication which is the operation of multiplying point *p* on an elliptic curve *E* defined over a field *GF(p)* with positive integer *k* which involves point addition and point doubling, spent 80% of execution time. Also the efficiency of operation *kP* is depends on the type of coordinate system used for point *p* on the Elliptic Curve and the algorithm used for recoding of integer *k* in scalar multiplication. An ample amount of analysis on binary fields and Elliptic Curve arithmetic for the NIST recommended elliptic curves was done in [4]. A dedicated implementation for the field GF ($2^{155}$) was done in [10]. More point multiplication algorithms can be found in [5] and [6]. Shantz [11] presented an efficient technique to calculate modular division, which is an vital arithmetic operation in ECC and other cryptographic system. Cohen et al [9] analyzed the impact of coordinate system in ECC implementation. They measured the performance of point Addition (PADD) and Point Doubling (PDBL) of different coordinate system. Malan et.al [13] implemented an ECC system using polynomial basis over binary field GF($2^m$).

So, aim of this paper is to propose a novel technique which can reduce the time required in scalar multiplication.

## III. INTRODUCTION TO ELLIPTIC CURVE CRYPTOGRAPHY

An elliptic curve $E$ over $GF(P)$ can be defined by

$$y^2 = x^3 + ax + b \text{ where } a,b \in GF(p) \quad (1)$$

$$4a^3 + 27b^2 \neq 0 \quad (2)$$

in the $GF(P)$. The point *(x, y)* on the curve satisfies the above equation and the point at infinity denoted by ∞ is said to be on the curve.

For example, $y^2 = x^3 + 2x + 5 \quad (3)$

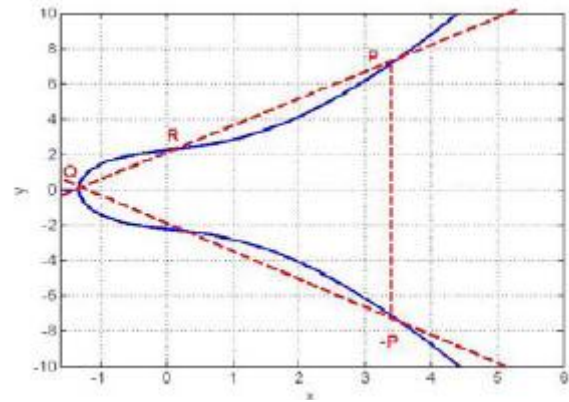

Figure 3. Elliptic Curve of equation (3)

$$y^2 = x^3 - 2x + 1 \quad (4)$$

The Elliptic Curve group operation is closed so that the addition of any two points is again a point of Elliptic Curve. Identity element of the group is point at infinity i.e. O and stratifies *P + O = O + P = P*.

For *P(x, y)* it can be define that *–P = (- x, y)* as the unique inverse of *P* in the group satisfying the property *P + (-P) = (-P) + P = O*.

A multiplication of a point *P* with an integer *k* can be expressed as the multiplication addition of one and the same point *k*-times.

$$Q = P + P + P + \ldots + P = k.P \quad (5)$$

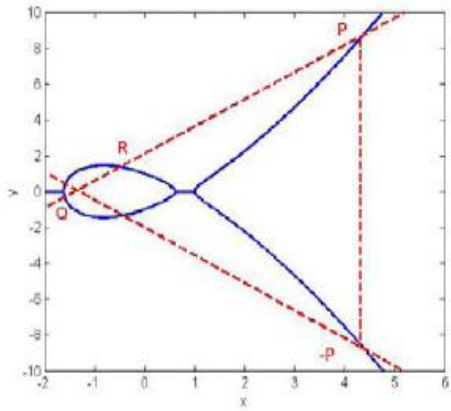

Figure 4. Elliptic Curve of equation (4)

The resultant product $Q$ is another point on the Elliptic curve. Given an Elliptic Curve $C$ over finite field $F_2^p$, a point $P$ and a product $Q$ the problem is to find a $k.N$ that holds $Q=k.P$. This problem is known as Elliptic Curve Discrete Logarithmic Problem (ECDLP) which is quite hard to solve. For example with a finite fields with $F_2^p$, $2^p$ elements takes about $O(2^{P/2})$ operations to find $k$ which is exponential amount of time. Thus the ECC depends on the complexity of the ECDLP, that is given points P and Q in the group, to find the number k such that Q=k.P. If there are two points on the curve namely, $P(x_1, y_1)$, $Q(x_2, y_2)$ and their sum is given by point $R(x_3, y_3)$ the algebraic formulas for point addition and point doubling are given by following equations:

$$x_3 = \lambda^2 - x_1 - x_2 \quad (6)$$

$$y_3 = \lambda(x_1 - x_3) - y_1 \quad (7)$$

$$\lambda = \frac{(y_2 - y_1)}{(x_2 - x_1)}, if \ P \neq Q \quad (8)$$

$$\lambda = \frac{(3x^2 + a)}{2y_1}, if \ P = Q \quad (9)$$

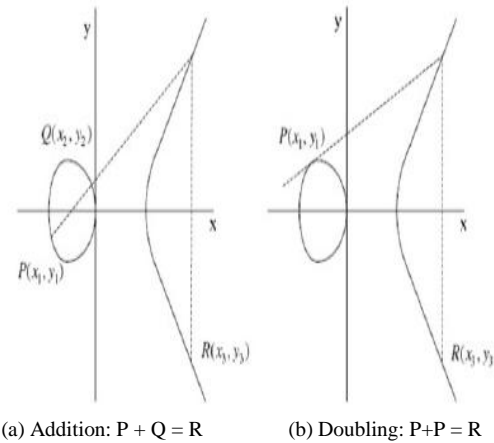

(a) Addition: P + Q = R     (b) Doubling: P+P = R

Figure 5. Elliptic Curve point addition and point doubling operations

IV. PROPOSED FLHSPKC TECHNIQUE

For reducing computational complexity of scalar multiplication proposed FLHSPKC techniques provides some novel schemes given in the following subsections.

*A. Hamming Weight Reduction Strategy*

The Hamming weight of a binary string is the number of symbols that are diverse from the "0" symbol. So in general Hamming weight is the number of "1" bits presence in the binary sequence. PADD is more computationally expensive than PDBL. Now, point multiplication can be decomposed to series of PADD and PDBL, because of computational complexness PDBL should be use more than PADD for point multiplication computational purpose. As an example consider $k= 511= (111111111)_2$. Now for any point P, 511P requires 8 point additions. However, if we write $(111111111)_2 = (1000000000)_2 - 1$. Then in decimal $511P = 512P - P = (1000000000)_2 P - P$ requires only one addition. Again point subtraction can be replaced by PADD because the inverse of an affine point $P(x, y)$ is $-P(x, -y)$. Thus this proposed strategy can implement in point multiplication to attain faster computation.

*B. Pairing Strategy*

Here pairing strategy for scalar multiplication in ECC has been used for computing $kP$ for any $k \in Z^+$ (set of positive integers). Let $P$ be a point on EC and let $k=3277$ then $3277P$ is equivalent to $(110011001101)_2 P$. Consider pairing size from 2 to 12; but it can be of any size. For a pairing size $PSIZE$, there are $(2^{PSIZE-1}-1)$ pre computations are needed i.e. if $PSIZE =3$ then there are, $(2^{3-1}-1) = 3$ pre computations, namely $3P$, $5P$ and $7P$. So increasing of pairing size also increase the number of pre computation and number of additions and doubling operations decreases. The numbers of PADD and PDBL for 3277P for different pairing size are as follows.

- Consider the pairing size $PSIZE = 2$ and $k= 763$ then $763 = (1011111011)_2$

    No of pre computations = $2^{PSIZE-1}-1 = 2^2 - 1 = [3] P$

    So, $763 = \underline{10}\ \underline{11}\ \underline{11}\ \underline{10}\ \underline{11}$

    The intermediate values of $Q$ are $P, 2P\ 4P, 8P, 11P, 22P, 44P, 47P, 94P, 95P, 190P, 380P, 760P, 763P$

    Computational cost = 9 doublings, 4 additions, and 3 pre computation.

- Consider the pairing size $PSIZE = 3$ and $k= 763$ then $763 = (1011111011)_2$

    No of pre computations = $2^{PSIZE-1}-1 = 2^3 - 1 = [7] P$

    The intermediate values of $Q$ are $5P, 10P, 20P, 40P, 47P, 94P, 188P, 376P, 381P, 762P, 763P$

    Computational cost = 7 doublings, 3 additions, and 7 pre computations.

- Consider the pairing size $PSIZE = 4$ and $k= 763$ then $763 = (1011111011)_2$

No of pre computations = $2^{PSIZE-1}-1 = 2^4 - 1 = [15] P$

The intermediate values of *Q* are *11P, 22P, 44P, 88P, 95P, 190P, 380P, 760P, 763P*

Computational cost = 6 doublings, 2 additions, and 15 pre computations.

- Consider the pairing size *PSIZE = 5* and *k= 763* then *763 = (1011111011)$_2$*

No of pre computations = $2^{PSIZE-1}-1 = 2^5 - 1 = [31] P$

The intermediate values of *Q* are *23P, 46P, 92P, 184P, 368P, 736P, 763P*

Computational cost = 5 doublings, 1 additions, and 31 pre computations.

- Consider the pairing size *PSIZE = 6* and *k= 763* then *763 = (1011111011)$_2$*

No of pre computations = $2^{PSIZE-1}-1 = 2^6 - 1 = [63] P$

The intermediate values of *Q* are *47P, 94P, 188P, 376P, 752P, 763P*

Computational cost = 4 doublings, 1 additions, and 63 pre computations.

- Consider the pairing size *PSIZE = 7* and *k= 763* then *763 = (1011111011)$_2$*

No of pre computations = $2^{PSIZE-1}-1 = 2^7 - 1 = [127] P$

The intermediate values of *Q* are

*95P, 190P, 380P, 760P, 763P*

Computational cost = 3 doublings, 1 additions, and 127 pre computations.

- Consider the pairing size *PSIZE = 8* and *k= 763* then *763 = (1011111011)$_2$*

No of pre computations = $2^{PSIZE-1}-1 = 2^8 - 1 = [255] P$

The intermediate values of *Q* are *95P, 190P, 380P, 760P, 763P*

Computational cost = 3 doublings, 1 additions, and 255 pre computations.

- Consider the pairing size *PSIZE = 9* and *k= 763* then *763 = (1011111011)$_2$*

No of pre computations = $2^{PSIZE-1}-1 = 2^9 - 1 = [511] P$

The intermediate values of *Q* are *381P, 762P, 763P*

Computational cost = 1 doublings, 1 additions, and 511 pre computations.

- Consider the pairing size *PSIZE = 10* and *k= 763* then *763 = (1011111011)$_2$*
No of pre computations = $2^{PSIZE-1}-1 = 2^{10} - 1 = [1023] P$
The intermediate values of *Q* are *763P*
Computational cost = 0 doublings, 0 additions, and 1023 pre computations.

*C. 1's Complement based Arithmatic Strategy*

In FLHSPKC scheme there exist another strategy based on 1's complement arithmetic. The formula used for calculating 1's complement is

$Comp = (2^{no.\ of\ bits} - 1) - N$ (9)

Where *Comp* = 1's complement of the binary number, *N* = respective binary number and

*no. of bits* = number of bits presence in binary form of *N*. To reduce number of intermediate operations of multiplication, squaring and inverse calculations used in ECC minimal non-zero bits in positive integer scalar are very important Now, equation (9) can be written as

$N = (2^{no.\ of\ bits} - 1 - Comp)$ (10)

For example, if *N =2046* then binary representation of *N* will be *N= (11111111110)$_2$*. Now calculate 1's complement of *N* i.e. *Comp = (00000000001)$_2$*. Number of bits present in binary representation of *N* is 11. After putting the respective values of *N, Comp* and *no. of bits* in equation (10) we get *2046 = ($2^{11}$ − 00000000001 − 1)*, this can be reduced as *2046 = (100000000000)$_2$− (00000000001)$_2$ −1*

So we have, 2046 = (100000000000)$_2$– ($2^1$− 0 − 1) −1 i.e. 2046 = 100000000000 − 1 −1 = 2048 − 1− 1.

The Hamming weight of *N* has reduced from 10 to 1 which will save 9 elliptic curve addition operations. For providing more optimized result above discussed 1's complement subtraction method is combined with sliding window strategy.

Now apply this proposed 1's complement arithmetic strategy to the same number 763 to show the effectiveness of algorithm with previously discussed proposed pairing strategy with *PSIZE=3*.

763 = (1011111011)$_2$

Using equation (10) we can write

763 = (10000000000)$_2$ – (0100000100)$_2$ -1

The intermediate values of Q are:

3P, 6P, 12P, 24P, 48P, 96P, 192P, 384P, 768P, 763P

Hence the Computational Cost = 8 doublings, 1 addition and 3 pre computations.

With *PSIZE*= 3, in pairing scheme for the same examples the intermediate values of *Q* are *5P, 10P, 20P, 40P, 47P, 94P, 188P, 376P, 381P, 762P, 763P and* Computational cost = 7 doublings, 3 additions, and 3 pre computations. So using 1's complement arithmetic scheme the computational cost has been reduced from 3 additions as in binary method to only 1 addition. The number of pre computations remained same. This can be proved for different *PSIZE*.

*D. Soft Computing based Arithmatic Strategy*

In this proposed soft computing based scheme fuzzy logic approach is used to perform recurring subtraction instead of performing division. For that some random multiples of m is used. This operation has certain fuzziness because the operation used is not exact but it involves some random multiples of m from the value to be reduced modulo m. The example of proposed fuzzy logic based scheme is shown in figure 6 and traditional binary multiplication of 2 binary integers is shown in figure 7. Let X=26 and Y=24 with m=17 then value of $X * Y \mod m$ will be 12. Since, fuzzy based approach mainly focuses on repeated subtraction instead of division for modulo reduction m. The process is shown in figure 6. The modulus m=17, a multiple of m is consider for speed of this process i.e. m.t=17*5=85.

- Whenever from number Y, "0" will be encountered then bit position weight from Y for corresponding "0" (find out bit position weight using 8-4-2-1 rules) will be multiplied by m.t value i.e. (Bit position weight x m.t) and this value will be subtracted.

- Whenever from number Y, "1" will be encountered then bit position weight from Y for corresponding "1" (find out bit position weight using 8-4-2-1 rules) will be multiplied by X value i.e. (Bit position weight x X) and this value will be added.

$X = (26)_{10} = (11010)_2$ and $Y = (24)_{10} = (11000)_2$

$$
\begin{array}{r}
(26)_{10} \quad (11010)_2 \\
\times (24)_{10} \quad (11000)_2 \\
\hline
-1 \times 85 \\
-2 \times 85 \\
-4 \times 85 \\
+8 \times 26 \\
+16 \times 26 \\
\hline
29
\end{array}
$$

$(29)_{10} \bmod (17)_{10} = (12)_{10}$

Figure 6. Soft Computing based modular multiplication

$X = (26)_{10} = (11010)_2$ and $Y = (24)_{10} = (11000)_2$

$$
\begin{array}{r}
11010 \\
\times 11000 \\
\hline
00000 \\
00000\times \\
00000\times \\
11010\times \\
11010\times \\
\hline
1001110000 = (624)_2
\end{array}
$$

$(1001110000)_2 \bmod (1001)_2 = (110)_2 = (12)_{10}$

Figure 7. Traditional modular multiplication

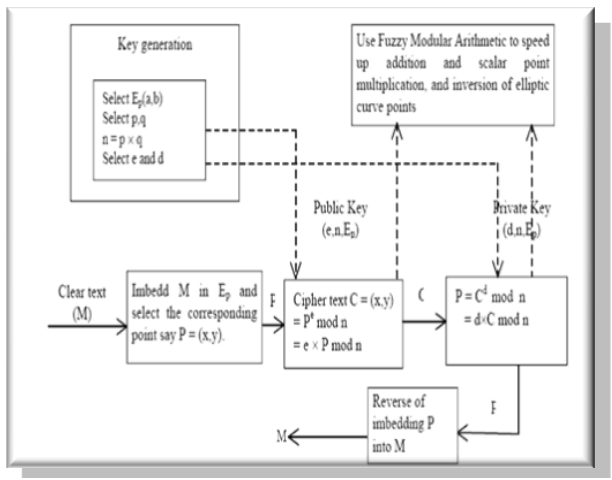

Figure 8. Soft Computing based Arithmatic Model

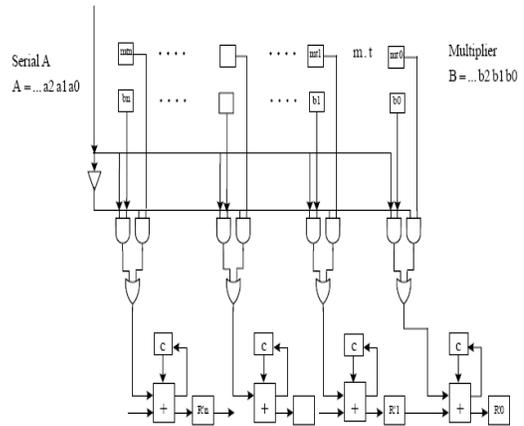

Figure 9. Array Multiplier for Soft Computing based Arithmatic

## V. PROPOSED SOFT COMPUTING BASED CONTROLLER FOR ECC

There is a tradeoff between the computational cost and the window size (*PSIZE* in proposed pairing scheme). Now this tradeoff is depends on balance between computing cost (or the RAM cost) and the pre computing (or the ROM cost) of the node. From this discussion it is noticed that the variety of wireless network working states will make this control complex and calculations could be relatively more expensive. Therefore, FLHSPKC ensures the optimum window size (*PSIZE* in proposed pairing scheme) is obtained by tradeoff between pre computation and computation cost with the help of fuzzy dynamic control system, to provide dynamic control. The goal of fuzzy decision problem is to maximize the minimum value of the membership functions for optimization purpose. Representation of multiobjective programming problem with the help of fuzzy optimization model is discussed follows.

$$Max : \min\{\mu_a(\phi)\} \,\&\, \min\{\mu_b(\varphi)\} \quad \forall a \in A \,\&\, \forall b \in B \quad (11)$$

Such that $\quad X_b \leq Y_l \qquad \forall b \in B \quad (12)$

$$\sum_{r \in R_p} z_{ra} = 1 \qquad \forall p \in P \,\&\, \forall a \in A \quad (13)$$

$$z_{ra} = 0 \text{ or } 1 \quad \forall r \in R \,\&\, \forall a \in A \quad (14)$$

Objective of the above equation is to maximize the minimum membership function of delays $\phi$ and $\varphi$ denotes the deviation value between recommend value and measured value.

Figure 8 shows a fuzzy control system with 3 inputs fuzzy controller. They are as follows.

- *Storage Room: it is the first input having* 3 states, namely (i) Min (ii) Intermediate (iii) Max
- *PreComputing:* The second input is pre-computing working load (*PreComputing*) in one of three states, namely (i) Min (ii) Intermediate (iii) Max
- *Doubling* : The third input is *Doubling*, expressing how much working load for the calculation "doubling" which has three states, namely (i) Min (ii) Intermediate (iii) Max

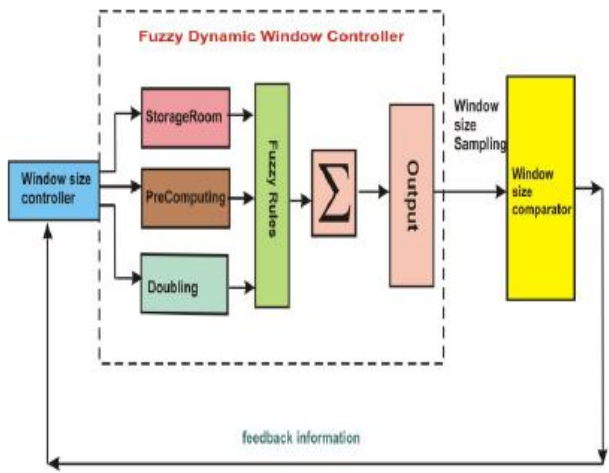

Figure 10. Proposed Soft Computing based Controller for ECC

The output is one, i.e.

- *WindowSize:* To express the next window size should be moved in which way, which has three states for the window sizes, namely (i) down (ii) stay (iii) up.

There are 26 *Fuzzy Rules* listed as shown in Table I where weights are unit.

- Mi=Min
- I=Intermediate
- Mx= Max
- D= down
- S= stay
- U=up

TABLE I.      SHOWS 26 FUZZY RULES

| StorgeRoom | PreComputing | Doubling | WindSize |
|---|---|---|---|
| Mi | Mi | Mi | U |
| Mi | Mi | I | U |
| Mi | Mi | Mx | S |
| Mi | I | Mi | U |
| Mi | I | I | U |
| Mi | I | Mx | S |
| Mi | Mx | L | U |
| Mi | Mx | I | S |
| Mi | Mx | Mx | S |
| I | Mi | Mi | U |
| I | Mi | A | U |
| I | Mi | Mx | S |
| I | I | L | U |
| I | I | I | S |
| I | I | Mx | D |
| I | Mx | I | S |
| I | Mx | Mx | S |
| Mx | Mi | Mi | S |
| Mx | Mi | I | S |
| Mx | Mi | Mx | D |
| Mx | I | Mi | S |
| Mx | I | I | S |
| Mx | I | Mx | D |
| Mx | Mx | Mi | D |
| Mx | Mx | I | D |
| Mx | Mx | Mx | D |

From the above table I it is observed that among 26 fuzzy rules only 9 fuzzy rules will be dominated the whole fuzzy control system. Considering only major 9 fuzzy rules quality of service (QOS) can be improved by decreasing the latency time of the system. These major 9 fuzzy rules are as follows:

1. *If (PreComputing is Min) and (Doubling is Min) then (WindowSize is Up)*

2. *If (PreComputing is Min) and (Doubling is Intermediate) then (WindowSize is Up)*

3. *If (PreComputing is Min) and (Doubling is Max) then (WindowSize is stay)*

4. *If (PreComputing is Intermediate) and (Doubling is Min) then (WindowSize is Up)*

5. *If (PreComputing is Intermediate) and (Doubling is intermediate) then (WindowSize is Up)*

6. *If (PreComputing is intermediate) and (Doubling is Max) then (WindowSize is stay)*

7. *If (PreComputing is Max) and (Doubling is Min) then (WindowSize is Up)*

8. *If (PreComputing is Max) and (Doubling is intermediate) then (WindowSize is stay)*

9. *If (PreComputing is Max) and (Doubling is Max) then (WindowSize is stay)*

In each above fuzzy conditions the value within a bracket denotes the weight number in unit. With 3 inputs StorageRoom, PreComputing and Doubling and 1 output i.e. WindowSize Mamdani composition rule base is used in each 9 fuzzy rules.

## VI.     RESULTS AND DISCUSSION

In this section results of each proposed novel strategy under FLHSPKC scheme has been presented in the following.

TABLE II.

PAIRING SIZE VS NO OF DOUBLINGS, ADDITIONS AND PRE COMPUTATIONS IN PAIRING STRATEGY

| PSIZE | Number of Doublings | Number of Additions | Number of Pre-computation |
|---|---|---|---|
| 2 | 9 | 4 | 3 |
| 3 | 7 | 3 | 7 |
| 4 | 6 | 2 | 15 |
| 5 | 5 | 1 | 31 |
| 6 | 4 | 1 | 63 |
| 7 | 3 | 1 | 127 |
| 8 | 3 | 1 | 255 |
| 9 | 1 | 1 | 511 |
| 10 | 0 | 0 | 1023 |

Figure 11 shows the tradeoff between window size and the computational costs in Pairing strategy.

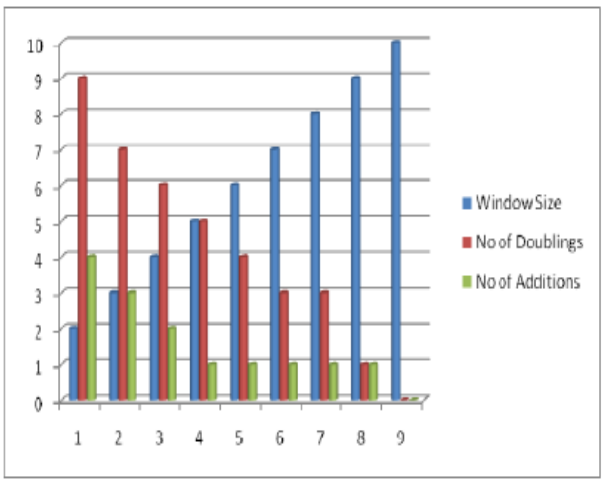

Figure 11. Graphical Representaion of Pairing Size and comutational cost in Pairing strategy

Figure 12 shows the tradeoff between window size and number of pre computations in Pairing strategy.

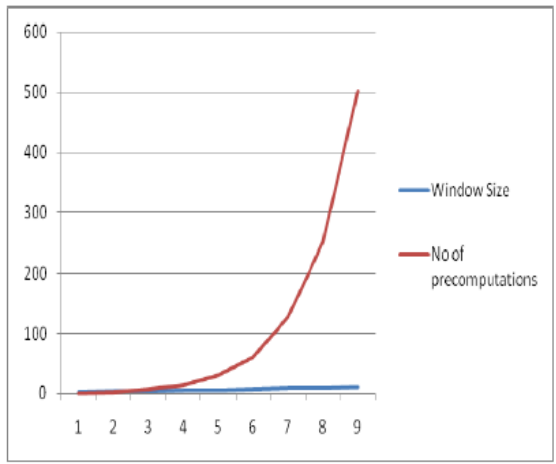

Figure 12. Graphical Representaion of Pairing Size and number of pre computation in Pairing strategy

TABLE III.

PAIRING SIZE VS NO OF DOUBLINGS, ADDITIONS AND PRE COMPUTATIONS IN 1'S COMPLEMENT BASED ARITHMETIC STRATEGY

| PSIZE | Number of Doublings | Number of Additions | Number of Pre-computation |
|---|---|---|---|
| 2 | 9 | 4 | 1 |
| 3 | 7 | 3 | 3 |
| 4 | 6 | 2 | 7 |
| 5 | 5 | 1 | 15 |
| 6 | 4 | 1 | 31 |
| 7 | 3 | 1 | 61 |
| 8 | 3 | 1 | 127 |
| 9 | 1 | 1 | 251 |
| 10 | 0 | 0 | 501 |

TABLE IV. EXPERIMENTAL RESULTS OF SOFT COMPUTING BASED ARITHMETIC STRATEGY (KB- KILO BYTES, MS- MILLE SECONDS, E-ENCRYPTION TIME, D-DECRYPTION TIME, AND T- TOTAL TIME)

| File size (KB) | ECC without Soft Computing based Arithmetic (Time in ms) | | | ECC with without Soft Computing based Arithmetic (Time in ms) | | |
|---|---|---|---|---|---|---|
| | *E* | *D* | *T* | *E* | *D* | *T* |
| 1 | 1490 | 1370 | 2860 | 1320 | 1340 | 2660 |
| 3 | 2470 | 2410 | 4880 | 1920 | 1980 | 3900 |
| 5 | 3510 | 3460 | 6970 | 2820 | 2890 | 5710 |
| 7 | 4573 | 4512 | 9085 | 3916 | 3996 | 7912 |
| 8 | 5685 | 5642 | 11325 | 4613 | 4663 | 9276 |

From table IV, it is observed that the time taken for encryption and decryption with fuzzy (soft computing strategy) is substantially decreased compared to corresponding counterpart without fuzzy.

Figure 13 shows the output with StorageRoom and PreComputing for fuzzy control in ECC scheme.

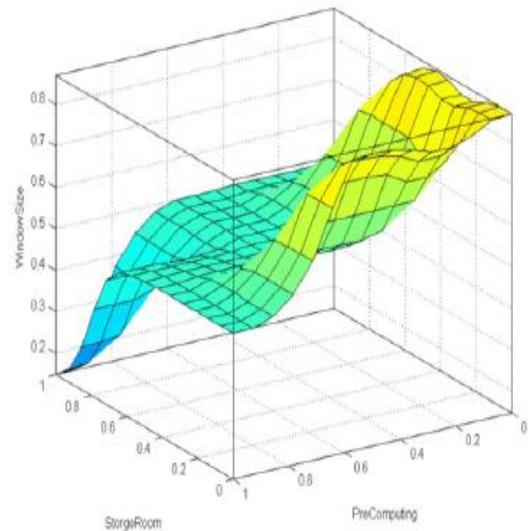

Figure 13. Graphical Representaion of output of the exterior StorageRoom vs. PreComputing for fuzzy control in ECC scheme

From the above figure 13 it is observed that in Min window size side *if the (StorageRoom is Min) the conquered function of "doubling" will play role*. But if the window size is at the Max side, the StorageRoom will be moderately stayed at the middle either for PreComputing or Doubling. Doubling will stridently increase when window size is little bit larger.

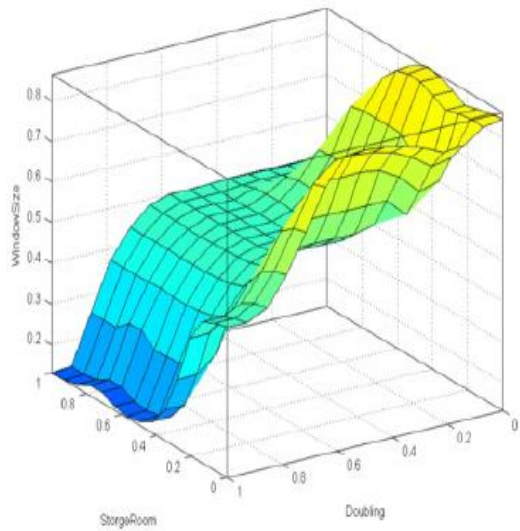

Figure 14. Graphical Representaion of output of the exterior StorageRoom vs. Doubling for fuzzy control in ECC scheme.

In figure 14 it is observed that larger window size for doubling is needed when StorageRoom is Max. Now set the weight value equals to 0.5 for the fuzzy rules 1, 5, 10, 13, 14, 15, 16, 18, 20, 21, 22, 23, 25, 26 then doubling will be increase by larger window size due to the fact that StorageRoom controlled the major functions. If PreComputing and Doubling are decreased by 0.02% and StorageRoom is increased by 0.04% then the output will change.

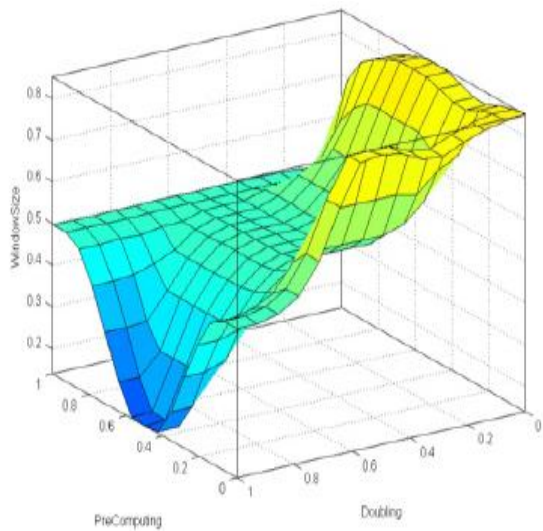

Figure 15. Graphical Representaion of output of the exterior StorageRoom =0.4 and PreComputing vs. Doubling for fuzzy control in ECC scheme

After initializing constant value to the StorageRoom i.e. StorageRoom = 0.4 the output of the PreComputing vs. Doubling is shown in figure 15.

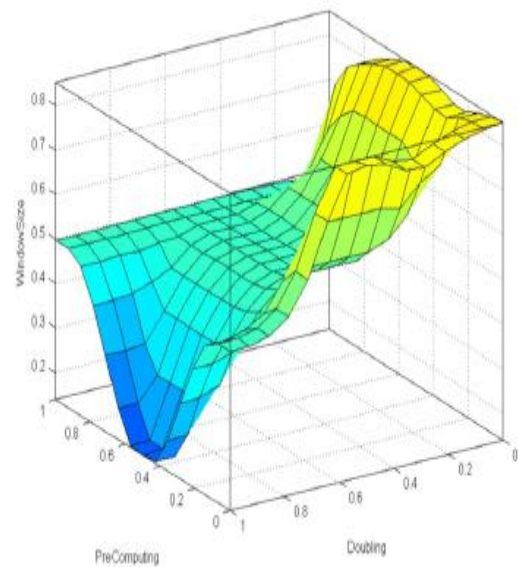

Figure 16. Graphical Representaion of output of the exterior StorageRoom =0.8 and PreComputing vs. Doubling for fuzzy control in ECC scheme

Figure 16 shows output of the PreComputing vs. Doubling with StorageRoom=0.8. This example shows that Storageroom is not dominating factor. The output of the "*PreComputing*" vs. "*Doubling*" is not much difference between Figures 8 and 9. It is noticed from figures 13 and 14 that the "dominated factor" is not the members of the "StorageRoom" in figure 15 due to the fact that the StorageRoom goes to large than 40% (or 0.4) the factors PreComputing and Doubling are dominated in figure 13 and figure 14 separately.

## VII. CONCLUSIONS AND FUTURE SCOPE

This paper presents a novel FLHSPKC technique which helps to provide a faster version of traditional ECC. In WSN ECC is the best alternative to RSA, DSA and other prime field based cryptographic algorithms. Because ECC offers same security level with fewer amounts of time and memory complexity compared to RSA, DSA. WSN has restraint in the bandwidth, computational potency, power availability or storage. Aim of this paper is to provide some alternative schemes which can be associated with traditional ECC to make it faster and suitable for WSN. Proposed FLHSPKC techniques provide following schemes a) Hamming weight reduction strategy, b) Pairing strategy, c) 1's complement arithmetic strategy, d) Soft Computing based Arithmetic Strategy.

Either of these strategies can be incorporated in traditional ECC for faster arithmetic computation. Soft computing based controller for ECC also been proposed in this paper. Pre-computing is related to the storage i.e. ROM and Computing is associated with the computing capability or capacity which is RAM. In this paper proposed soft computing based controller always controlled the window size which is a trade off between available ROM and RAM in a sensor node for a particular instance of time.

Future scope of this FLHSPKC technique is that

- FLHSPKC technique deals with only integer set. So, special character constant can also be considered with integer set for better security purpose.
- For security reason key length of the proposed method can also be increased.

ACKNOWLEDGMENT

The author expressed deep sense of gratitude to the Department of Science & Technology (DST), Govt. of India, for financial assistance through INSPIRE Fellowship leading for a PhD work under which this work has been carried out, at the department of Computer Science & Engineering, University of Kalyani.

AUTHORS PROFILE

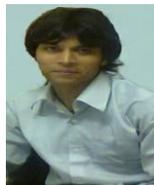

**Arindam Sarkar**

INSPIRE Fellow (DST, Govt. of India) at the department of Computer Science & Engineering, University of Kalyani, MCA (VISVA BHARATI, Santiniketan, University First Class First Rank Holder), M.Tech (CSE, K.U, University First Class First Rank Holder). Total number of publications 14.

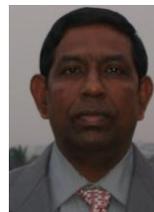

**Jyotsna Kumar Mandal**

M. Tech.(Computer Science, University of Calcutta), Ph.D.(Engg., Jadavpur University) in the field of Data Compression and Error Correction Techniques, Professor in Computer Science and Engineering, University of Kalyani, India. Life Member of Computer Society of India since 1992 and life member of cryptology Research Society of India. Dean Faculty of Engineering, Technology & Management, working in the field of Network Security, Steganography, Remote Sensing & GIS Application, Image Processing. 25 years of teaching and research experiences. Eight Scholars awarded Ph.D. one submitted and 8 are pursuing. Total number of publications 252.